# The Impact of Social Media in Learning and Teaching: A Bibliometric-based Citation Analysis

**A. K. Shaikh, S. Ali and R. Al-Maamari**

Department of Information Systems, Sultan Qaboos University, Muscat, Oman,
shaikh@squ.edu.om, saqib@squ.edu.om, u111514@student.squ.edu.om

**Abstract –** This paper presents the results of a systematic review of the literature on the impact of social media in learning and teaching through bibliometric-based Citation analysis. The objective of the review was to map the evolution of the current literature and identify the leading sources of knowledge in terms of the most influential journals, authors, and articles. From a total of 50 top most relevant articles, selected from the Scopus database, a detailed citation analysis was conducted. The study explored the overall theoretical foundation of social media research involving in learning and studying; and identified the leading sources of knowledge in terms of the most influential journals, authors, and papers; and revealed research trends over the last four years by citation analysis. The analysis of citation data showed that "International Journal of Management Education" is the leading journal in social media in learning and teaching research. Author *"Abdullah Z."* was found to be the leading author in this field in terms of a total number of publications, total citations, and h-index, while the most cited article was authored by "Baaran S." and by "Bapitha L.". The contribution of this study is to clearly outline the current state of knowledge regarding social media in learning and teaching services in the literature.

---

* Corresponding Author: Abdul K. Shaikh, shaikh@squ.edu.om

## Introduction

Internet and social media have a significant impact on our daily life including our way of learning and teaching. According to the latest smart insight report[†], more than 53% of the global populations are now connected via social media. Social media can characterize as PC-interceded advancements that help the client to make and share information, data, and different things through the system and virtual correspondence [1]. To take the advantages of existing social media network and platforms, the teaching and learning activities can be enhanced. This could be an opportunity c for colleges and schools to apply social media in learning and teaching. To fulfil this gap, in this paper, we use a bibliometric technique [19] for comprehensive analysis of social media-related articles, which have involvement in learning and studying in centralized control [20]. Bibliometric incorporates various techniques for the examination of data, for example, bibliographic coupling utilizing citation, citation analysis, content analysis, and co-word examination utilizing keywords [2, 3]. In this study, we use citation analysis. The bibliometric investigation gives an understanding of the development of writing and the progression of information inside a predefined field over some time by breaking down the data accumulated in the database, for example, references, writers, keywords, or the scope of journals [3].

The rest of the sections are outlined as: literature review section explores the overall theoretical foundation of social media research involving learning and studying including research questions and its motivation, followed by the section on research methodology and discussion section on the bibliometric results. The last section presents the conclusion and the possible future research.

## Literature review

The literature review research method includes three actions: "planning, execution, and reporting". Each of these actions are partitioned into further sub-actions. Planning incorporates separating the remaining task to complete on a time and building up the "retrieve protocol itself". The planning action combines study questions and building up the research technique, the consideration/avoidance measures, and the information extraction structure. The execution action incorporates information recovery, study choice, information extraction, and information blend. At last, the revealing actions present and decipher the outcomes.

There are numerous collections of research periodicals accessible to look over in both electronic (online) and physical (hard copy) format structure. This research limited to extract peer-reviewed journals from electronic (online) format available. The minimum criteria to include SCOPUS indexed periodicals for this study.

---

[†] https://www.smartinsights.com/social-media-marketing/social-media-strategy/new-global-social-media-research/

## *Research questions and motivation*

In this study, three research questions arise which help to clarify and motivate the contribution of social media in learning and teaching through citation analysis.

RQ1. What is the impact of social media in teaching and learning which is investigated by researchers?

RQ2. What application of social media is using more among students?

RQ3. What is the social media involvement in learning and studying research goals?

Research question 1, encourages the discovery of the different impacts of social media in learning and studying either positive or negative. For the second question, the main drive is to know which applications are more interesting for the students to use in education. Further, the main inspiration of the third question is to figure out the vast interest of research in this field and which areas might be under-contemplated: investigating fundamental ideas, gathering information on current practices, or targeting propelling practice through science plan. To answer the given research questions, 50 papers are classified on the work by Piattini, Genero, Poels, and Nelson [4].

### *Q1. What is the impact of social media on teaching and learning either positively or negatively?*

Social media refers to a wide range of applications in creating, sharing, commenting, and discussing digital content. Other aspects of social media that are often overlooked is the ability to transform education into a more social, open, and cooperative place [5]. Consistent with theoretical perspectives that are related to social media, there are many studies and research that are providing empirical evidence on the impact of social media in learning and teaching for students [6]. There are a large number of studies in this field, that believe and agree that the use of social media has many benefits for students [7-9],[10]. Such as increasing students' collaboration, participation, forming a team projects [7], sharing the most important information, and asking questions by using social media platforms [11]. Social media enhances studying, learning and increases studying demand. Some benefits are listed below:

**Knowledge construction:** In this aspect, we found many processes which are more active such as, ask questions, exchange ideas, give opinions, share information and resources. This knowledge construction is more responsible for the conversion process in the social media environment.

**Knowledge sharing:** Teachers are using social media such as Facebook for sharing knowledge and communicating with other people of similar thinking and reaching multiple audiences.

**Flexibility and accessibility:** That means you can learn at any time and place from the applications of social media. As many teachers and students use Twitter because it is easier to use and easier to access [12].

**Students want virtual office hours with their teachers:** From the study found in 2013, students appreciate the opportunity to make virtual discussions with their teachers during office hours to support their learning and increase flexible communication with their teachers.

**Many students are fluent in social media:** Studies [3] revealed the use of the internet between teenagers i.e., 77% of users are between 12 and 17 years, and 87% of users between the age of 18 and 29. Also, it found that most users are using Facebook applications for their learning

On the other hand, some studies are explaining the negative side of social media in teaching and studying [13-15]. For example, many of the bloggers and writers post wrong and inappropriate information on social media applications which leads the education system to failure [13]. For more, most students misuse social media by creating fake accounts for entertainment, and after that, they become addicted to it [11]. Some students aren't completely aware of the policies and terms that social platforms are already using [11, 14]. Laura Rueda, Jose Benitez, and Jessica Braojos showed that social media can also affect teaching activities, while some students dislike sharing their ideas with their teachers through social media and face some distractions like games, notifications about the video, advertisements, etc. [15].

### Q2. What are the applications of social media used by students?

There are some studies that showed how social media tools can impact learning and teaching [5, 16]. It can improve the performance of students in their studying, increase their knowledge, easy ways for understanding their lessons [9], and a more active way to connect with other students and teachers [16]. However, from the literature review, other social media-enabled tools are explored. For example, Facebook is a popular application among people, it can help students to generate information, references, and make a group for students. Another popular application of social media is Instagram and Twitter, which allow teachers and students to communicate and collaborate [17]. The students also can be satisfied when using YouTube to see different videos about their lectures than traditional ways in the classroom [5].

### Q3. What is the social media involvement in learning and studying research goals?

The purpose of investigating the goals of the research papers is to determine where social media involvement in studying and learning research interest lies and to determine which areas may be under-studied. There are 18 papers (which represents 36% of the total) related to assuring the impact, 13 papers (26%) are related to evaluating the impact, and 10 papers (20%) to measure the impact. The other two categories, improving and understanding, together account for less than 10% of the papers. Research on applied areas of the social media involvement in learning and studying assurance techniques and the evaluation of the impact of social media in learning and studying account for more than three-quarters (67.33%) of the papers published. This is not surprising, as the impact of social media on learning and studying is a critically important topic. This means that a given topic is understudied or that it has yet to find acceptance with journals, or it may be that both of these states are the case for the topic in question.

## Research methodology

In this section, the research methodology is explained where the research selection process and citation analysis were presented.

### *Research selection process*

The first step in this section was to select papers from the Scopus database. The papers that are selected are written in the English language. The document type was determined as "journal articles" to maintain the quality. The keywords selected were "Social media for studying" and "social media for sharing information" in the topic section, using the advanced search function. In the first search, a total of 188 articles relating to the use of social media involvement in learning and studying were selected. Then, the study identified the relevancy of the topic whether it is relevant to social media learning or teaching or not. Consequently, by excluding the irrelevant papers, 50 articles were selected for analysis from different SCOPUS indexed peer-reviewed journals. The citation analysis, therefore, covers published articles spanning the period 2017 to 2020.

**Citation Analysis:** It is a fact that there is a huge number of papers in every field. So, it became very difficult to understand and analyze the number of papers using traditional tools. To address this challenge, the use of bibliometric analysis is a vital technique. Citation analysis has become a very important and useful tool that provides a systematic analysis of a large number of earlier literature over time, comparing the relative impact of most prolific authors, top journals, and highly cited papers [18].

## Results and discussion

The results of citation analysis are demonstrated into three categories: (1) the timeline of citation, (2) the most effective papers in the field of social media involvement in learning and teaching, and (3) the most influential and qualified authors.

**The timeline of citation and publication:** Figure 1 shows the average article citations per year from 2017 to 2020. From Figure 1, in 2017 and 2018 there have been many citations during these years. For more, in 2017 the average citations are 2.889 per year. In 2018 the average number of citations is 4.35786 per year. The average citation in 2019 was 4.35786 per year. But no records of citations in 2020.

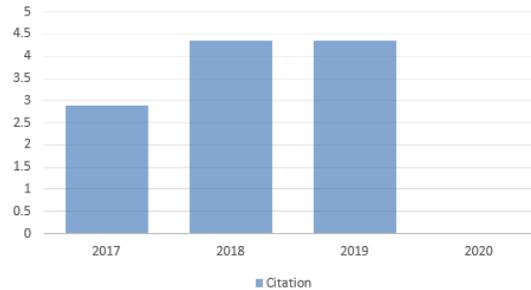

**Fig. 1** Average Article Citations per Year (2017-2020)

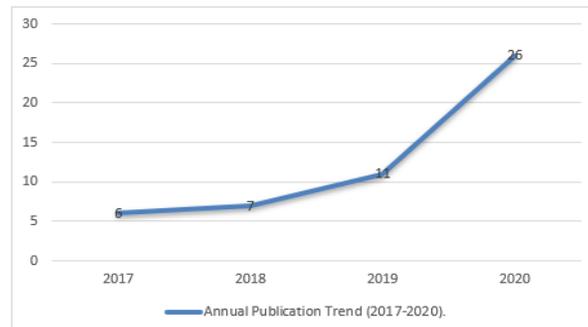

**Fig. 2** Annual Publication Trend (2017-2020)

Figure 2 shows the annual publication trend from 2017 to 2020. According to Figure 2, the most articles published were in 2020, which means this issue has been increasing more among the students in the particular year. In 2020, the annual publication was 27 studies, whereas, 2019 about 11 studies, seven studies in 2018, and six studies in 2017 were conducted respectively.

**The most effective papers in the field of a social media impact learning and studying:** Many journal papers are written in the field of social media on learning and studying efficiency. However, we identified the 8 most effective papers based on the total number of citations, which are presented in Table 1. According to this, the top five papers are "Social media as a complementary learning tool for teaching and learning: the case of YouTube", "Students wellbeing, fear of missing out and social media engagement for leisure in higher education learning environments", "Networked scholarship, and motivations for social media use in scholarly communication", "Social teaching: student perspectives on the inclusion of social media in higher education", and "Uses and gratifications factors for social media use in teaching: instructors perspectives". These top 8 papers account for 15.32% of the total citations until 2020.

Social media as a complementary learning tool for teaching and learning: the case of YouTube was the top-ranked paper, publishing 0.6% of the total number of papers (50 papers). This paper were cited 26 times, which corresponds to 5.6% of the total citations. A potential issue with the comparison of total citations is that the newest papers have less, or no citations compared to older papers. Likewise, some journals have published more papers on a particular issue than other journals, and, as a result, they have accumulated more citations. To account for this issue, another way of comparing the journal rankings is to look at the average citations per year. The average citations per year function take out the impact of "years since publication" and "number of papers". These papers received 6, 7, and 17 citations, respectively. The calculation of average citations per year for this journal is shown in equation 1,

Per year citations (paper 1) = Total Citations / (2020 – 2017)  = 6 / 3 = 2 (1)

Based on average per year citations, the top five journals have on average 8.667, 6.000, 4.250, 4.250, and 5.000 citations, respectively.

**The most influential and qualified authors:** This section will provide the most impact and productive authors in the field of social media involvement in learning and studying. For that, table 1 shows the top 8 most influential authors in order, which is based on total citations. Also, it provides more detail about some metrics such as g-index, h-index, m-index, and the number of publications to make comparisons between the contribution and influence of each author. The top five authors based on total citations are Abdullah Z, Adeniyi E.A., Agllias K., Aitchison C., and Ajamu G.J. While, based on h-index, g-index, and m-index, approximately all authors on the same level. The greatest number of citations done by Baaran S. is 26 and by Bapitha L. are also 26 citations.

This section discusses the results and comparative analysis of the total sample of 50 articles on social media involvement in studying and learning. The study is divided into five categories which are: (1) a summary of those characteristics in a geographical, (2) time evolution of the published work in the field, (3) identification of the most influential journal in the field, (4) identification of the most prolific author, and (5) identification of the most influential paper in the field.

Basic summary of the selected articles: According to a geographic point of view, there are many countries like North Korea, the USA, Nigeria, Malaysia, and Africa that have shown more contribution in social media research than others. The one main reason is, these countries have many advanced universities in technology. Also, many authors are interested in this field because they know while using social media in university or school has a lot of impact on learning and studying for the students either positively or negatively.

**Table 1** Top 8 Most effective papers in the field of how social media impact studying and teaching.

| Paper Title | Publication | | Total Citations | | Average Citations per Year (APY) | | |
|---|---|---|---|---|---|---|---|
| | Number | % | Number | % | Rank1 | Number | Rank 2 |
| The use of social media (some) as a learning tool in healthcare education: an integrative review of the literature. | 86 | 4.3% | 0 | 0.00% | 12 | 0.000 | 12 |
| Social media as a complementary learning tool for teaching and learning: the case of YouTube. | 79 | 3.95% | 26 | 3.6% | 1 | 8.667 | 1 |
| Students' wellbeing, fear of missing out, and social media engagement for leisure in higher education learning environments. | 51 | 2.55% | 18 | 2.56% | 2 | 6.000 | 6 |
| The best in disaster project: analyzing and understanding meaningful YouTube content, dialogue, and commitment as part of responsible management education. | 34 | 1.7% | 0 | 0.00% | 13 | 0 | 13 |
| Social media for teaching and learning within higher education institution: a bibliometric analysis of the literature (2008-2018). | 22 | 1.1% | 15 | 1.60% | 3 | 5.000 | 2 |
| Social teaching: student perspectives on the inclusion of social media in higher education. | 16 | 0.8% | 17 | 1.55% | 4 | 4.250 | 4 |
| Networked scholarship and motivations for social media use in scholarly communication. | 13 | 0.65% | 17 | 1.52% | 5 | 4.250 | 3 |

| | | | | | | |
|---|---|---|---|---|---|---|
| Social media as a tool for extending academic teaching & learning or influence of social media in students' academic performance. | 4 | 0.2% | 3 | 0.44% | 6 | 0.75 | 7 |

**The time evolution of the publications and citations in the social media involvement in learning and studying:** the development of using social media in the study and learning has led authors to write about the influence of social media on the students. Another reason for the huge increase in the development of articles in this field is the large number of people who have a mobile phone, which led to an increase in the number of accounts in social media for discussion among students.

Figure 3 shows the evolution of the average article citations per year, in the social media involved in the learning and studying field. For more, in 2017 the average citations are 2.889 per year. In 2018 the average number of citations is 4.35786 per year. The average citation in 2019 was 4.35786 per year. But no records of citations in 2020.

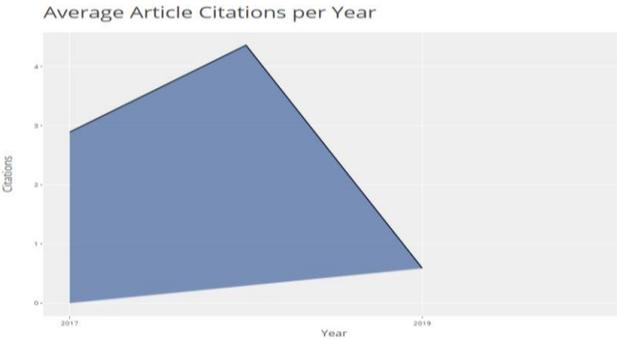

**Fig. 3** Annual average article citations in the social media involving in learning and studying

It is evident that older papers are expected to have received more citations than the more recent ones which have not been read by many readers yet. That can explain why the articles in 2020 have no citations yet.

This study has performed a measure of journals in four ways i.e., ranking based on total citations, based on total global and local citations ranked, and ranking based on the h-index. The study provided with the following information about the six metrics that the top-10 journals in social media involved in the learning and studying field that are used in this research. For that, none of the journals manages to achieve its ranking for all 6 metrics. Only one journal has managed to retain the same ranking in five out of six metrics. Which is the "International Review of Research in Open and Distance Learning".

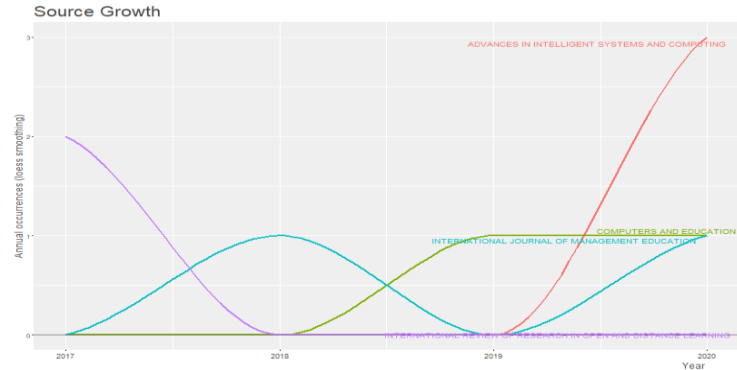
**Fig. 4** Cumulative occurrence of social media involvement in publications

The cumulative growth in the annual publication of social media involvement in learning and studying papers in the top four journals concerning their h-index is shown in figure 4. From figure 4, the "Computers and Education" journal was from 2017 to 2018 it was suspended and after 2018 it increased to 2019 in research production. The journal "International Journal of Management Education" has oscillating between high and low, which is 2018, and 2020 has the most time for production in this field. While "Advances in Intelligent Systems and Computing" have increased significantly from 2019 to 2020. The journal "International Review of Research Open and Distance Learning" has a lot of articles published in 2017, but after 2017 the publication has stopped in this field.

**Most prolific author:** Paper writing productivity of each social media involvement in learning and studying authors is measured in four ways: the first measure is the number of publications in which one's name appears as an author. However, since one author's contribution is reduced when more authors are added to the paper, we also apply a second measure which counts the number of fractional publications. This number adjusts productivity with the number of authors in one paper. The third measure is the total citations of one author in the social media involvement in the learning and studying field and the fourth measure is one's h-index on this specific field. Since author productivity ranking is described with four metrics, it is important that for an author to establish a definite ranking, one must have the same ranking in all four metrics, or a higher one in one of the metrics. For example, if an author gains the second position in three out of the four metrics, but in the fourth metric he has gained a higher position than in the 2$^{nd}$, namely he has gained the 1$^{st}$ position, then we can speak definitely about a 2$^{nd}$ position. Overall, we observe two authors that have a strong ranking position in three out of the four metrics. These authors are *"Cooke S."* and "Abdullah Z.".

**Table 2** Ranking of authors in different metrics

| Author | h-index | g-index | m-index | TC | NP | PY_Start |
|---|---|---|---|---|---|---|
| ABDULLAH Z | 1 | 1 | 0.250 | 3 | 1 | 2017 |
| ADENIYI EA | 0 | 0 | 0.000 | 0 | 1 | 2020 |
| AGLLIAS K | 0 | 0 | 0.00 | 0 | 1 | 2020 |
| AITCHISON C | 0 | 0 | 0.000 | 0 | 1 | 2020 |
| AJAMU GJ | 0 | 0 | 0.000 | 0 | 1 | 2020 |
| AL-RAHMI WM | 0 | 0 | 0.000 | 0 | 0 | - |
| ALBERTH A | 0 | 0 | 0.000 | 0 | 1 | 2020 |
| ALSHAYA H | 1 | 1 | 0.333 | 1 | 1 | 2018 |
| ALSHEHRI O | 0 | 0 | 0.000 | 0 | 1 | 2019 |
| ALT D | 1 | 1 | 0.333 | 18 | 1 | 2018 |
| ARIS B | 1 | 1 | 0.250 | 3 | 1 | 2017 |
| ATALLA S | 1 | 1 | 0.333 | 1 | 1 | 2018 |
| AWOTUNDE JB | 0 | 0 | 0.000 | 0 | 1 | 2020 |
| AYO FE | 0 | 0 | 0.000 | 0 | 1 | 2020 |
| BAARAN S | 0 | 0 | 0.000 | 0 | 1 | 2019 |
| BALOUL MS | 0 | 0 | 0.000 | 0 | 1 | 2019 |
| BAPITHA L | 0 | 0 | 0.000 | 0 | 1 | 2020 |
| BERGDAHL N | 1 | 0 | 0.000 | 0 | 1 | 2020 |
| BLAKEMORE T | 1 | 0 | 0.000 | 0 | 1 | 2020 |
| BOLD U | 1 | 0 | 0.000 | 0 | 1 | 2019 |

Table 2 provides the ranking of additional social media involvement in learning and studying authors and their ranking is based on additional metrics. It ranks 20 authors in six metrics. The h-index and the g-index offer the same ranking to all authors and this is expected and explained by the nature and the definition of the two indexes. The m-index, however, provides a completely different ranking. Only *"Abdullah Z."* keeps the same first position in the h-index, g-index, and NP (Number of papers). The different ranking also continues with total citations. Table 2, can have different ranks for authors with these different metrics.

The paper presents ten origin countries from where publications in social media involvement in learning and studying have stemmed the most. The origin of the publication does not signify the country origin of the author, but the country the author's affiliation. Thus, the figure identifies ten major origin countries from where publications in the social media involvement in learning and studying have stemmed. Once again it is verified something we have already mentioned, i.e., that "American Countries" are on the rise in social media and therefore they are keener to investigate the topic. Two Asian countries appear as well in this figure. These are *"Malaysia"*, and *"Indonesia"*.

The highest number of studies originates from affiliations in "USA", then "Malaysia" by publishing 3 articles from 50 articles. "Indonesia" and "Sweden" have published two articles in each country in this study. Other remaining countries have published one article in this study.

Moreover, the "social media" keyword specifically is expected to follow the same growth trend as the number of publications themselves. According to Figure 5, there is a clear upward tendency in the keyword "social media" which is due to the usage of this keyword in almost all papers. The rest of the lines reveal a less steep evolution of the keywords. This occurs in "students", "social networking online", "human experiment", "teaching". The rest of the keywords such as "education", "WhatsApp, "high education", occur at a constant level at all times up to date.

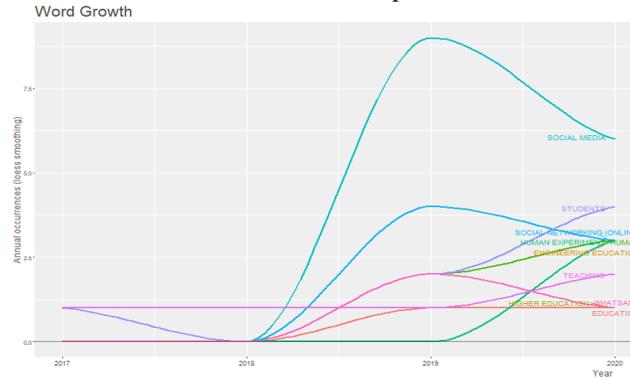

**Fig. 5** Evolution of different keywords

## Conclusion

Social media has had a significant impact on individual lives, including our way of teaching and learning. The motivation to conduct this study arose from the fact that to identify the potential impact of social media on learning and teaching and which social media application is most commonly used for teaching and learning from the student's perspective. The researchers addressed a comprehensive selection of studies published in Scopus-indexed journals between 2017 and 2020, inclusive. The systematic review uncovered 50 articles relevant to learning and teaching via social media, which were analyzed in terms of "most influential paper," "key authors," and "keywords.      and revealed research trends over the last four years by citation analysis. This will be beneficial for researchers and practitioners who require a reliable knowledge base from which to launch further research into learning and teaching via social media. Future work can be carried out to conduct content analysis to evaluate and identify trends and themes in data sets and identify self-citations of authors that may some impact on research quality.

# Biography

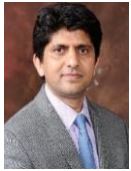

**Abdul Khalique Shaikh is** an Assistant Professor associated with the Department of Information Systems at Sultan Qaboos University Muscat Oman. He has received his PhD degree from a highly reputable Australian University Monash in 2013 and Master degree from University of Detroit USA with distinction. His research interest includes Social Network Analytics, Big Data Analytics, Data Science, Data Governance, E-participation and Blockchain Technology.

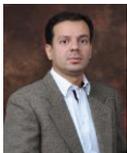

**Saqib Ali** is a business, IT implementation and execution professional, with strong skills in business analysis, Software development, management, and implementation, developed through higher education and useful work experience. Strong understanding of developing business strategies, business process integration and their alignment with IT, along with innovative practices to gain or sustain a competitive advantage. Dr. Saqib has also been active in teaching and R&D (Research and Development), he successfully able to attract number of internal, external and strategic grants while working at SQU.

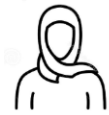

**Ramla Al-Maamari** is a final year undergraduate student in the Department of Information Systems, College of Economics & Political Science at Sultan Qaboos University. Her research interests include E-learning, E-commerce, Social Media and Technology Adoption Model.